\documentclass[aps,pre,twocolumn,showpacs,floatfix,superscriptaddress]{revtex4}
 \usepackage{graphicx}
\usepackage{epsfig}

\begin{document}

\title{On Dynamics and Optimal Number of Replicas in Parallel Tempering Simulations} 

\author{Walter Nadler}
\email{wnadler@mtu.edu}
\affiliation{Department of Physics, Michigan Technological University, Houghton, Michigan, USA}

\author{Ulrich H. E. Hansmann}
\email{hansmann@mtu.edu, u.hansmann@fz-juelich.de}
\affiliation{Department of Physics, Michigan Technological University, Houghton, Michigan, USA}
\affiliation{John-von-Neumann Institute for Computing, 
             Forschungszentrum J\"ulich, D-52425 J\"ulich, Germany}

\date{\today}

\begin{abstract}

We study the dynamics of parallel tempering simulations, also known as the replica exchange technique, which   has become the method of choice for simulation of proteins and other complex systems.
Recent results for the optimal choice of the control parameter discretization allow a 
treatment independent of the system in question. 
Analyzing mean first passage times across control parameter space, we find 
an expression for the optimal number of replicas in simulations covering a given temperature range.
Our results suggest a particular protocol to optimize the number of replicas in actual simulations.

\end{abstract}

\maketitle

The effective simulation of proteins, glasses and similar complex systems has
remained one of the defining challenges in computational physics. 
The main problem in  such simulations is slow relaxation due to barriers 
and bottlenecks.  Parallel tempering  - also known as the replica exchange 
method -  promised a way out of this dilemma~\cite{Geyer1995,HN,H97f}. 
Here, canonical or generalized-ensemble  simulations~\cite{HO96g} 
are performed in parallel  at  different values of a control parameter,  most often the temperature.
At certain times the current conformations of replicas at neighboring control parameter values are exchanged according to a generalized Metropolis rule~\cite{Metropolis}. 
An individual replica  performs a random walk in control parameter space,
allowing it to enter and escape local free energy minima. As a consequence, the state space is explored more evenly, especially e.g. at low temperatures.

Replica exchange simulations are usually performed on massively parallel machines, 
with one or more computing nodes dedicated to performing the simulation of a replica at a particular control parameter value. 
In order to optimize the use of resources, an important  question is whether 
an optimal choice for the number of replicas, 
--- or, equivalently, control parameter values~\cite{footnote1} --- exists, and how 
it might be determined. 
To our knowledge, no systematic investigation of this particular problem has been done to date.  

A likely reason is that this question is closely connected to the meta-dynamics of parallel tempering,
and a  full understanding of this method for complex systems --- exhibiting broken ergodicity~\cite{Palmer}--- is still missing~\cite{NH2007}.  It is also not a well-posed problem.
Apart from the number of replicas, the main adjustable parameter is the $distribution$ of control parameter values. 
The optimal number of replicas will depend strongly on the strategy used for the 
temperature  discretization. Usually,  a constant discretization is employed. 
However, often bottlenecks exist in some control parameter regions, 
and speeding up the equilibration of the system 
is possible only by using a finer discretization in these regions.
In order to investigate the question of the optimal number of replicas systematically, 
a method of discretization has to be employed that also complies with some optimality criteria. 

Major advances have been made recently in that direction.
Instead of concentrating on stationary distributions that arise from sampling, 
Trebst et al.~\cite{Trebst2004,Trebst2006a} have investigated the  $flow$ across control parameter space and have 
provided an iterative scheme for adjusting the discretization to optimize the flow distribution. 
Subsequently, we have shown that optimizing the flow is equivalent to minimizing the total first passage time to cross control parameter space~\cite{NH2007}.
In this letter, we  use those previous results as a basis for investigating the optimal number of replicas. 
We restrict ourselves to the situation of optimized flow and determine which number of control parameter values --- identical to the number of replicas --- minimizes the first passage time of a single replica.

In the following, we will consider parallel tempering with $N+1$ replicas. Hence, we will assume $N+1$ different control parameter values $\beta_0<\beta_1<\ldots<\beta_N$, i.e. we have $N$ control parameter intervals $\left[\beta_n,\beta_{n+1}\right]$.
We will also use the conventions $\beta_0=\beta_{min}$ and $\beta_N=\beta_{max}$. For simplicity we will call the control parameter a (inverse) temperature in the rest of this paper.

The time evolution of the probability $P(n,t)$ that an individual replica is 
at temperature $\beta_n$ at time $t$ 
can be approximated by a Master equation~\cite{Gardiner} in discrete time~\cite{NH2007}
\begin{eqnarray} 
P(n,t+1) &=& P(n,t)\times 
\nonumber  \\ 
& & \left[1 - W(\beta_n\to \beta_{n-1}) - W(\beta_n\to\beta_{n+1})\right] +
\nonumber  \\ 
& & P({n-1},t)W(\beta_{n-1}\to \beta_n) + 
\nonumber  \\ 
& & P({n+1},t)W(\beta_{n+1}\to \beta_n)  \quad , 
\label{eq:MET}
\end{eqnarray}
where $W(\beta\to\beta')$ are transition probabilities between neighboring temperatures. Of  course, these probabilities depend on those temperatures, 
and the master equation for replica exchange  is 
characterized by symmetric transition probabilities, 
\begin{equation}
W(\beta\to\beta') = W(\beta'\to\beta)  
\equiv W(\beta,\beta') =  W(\beta',\beta) \quad .
\label{eq:W}
\end{equation}
The relation between stationary flow $J$ and first passage time $\tau$ in one-dimensional stochastic systems has been investigated in the context of channel flow in Ref.~\onlinecite{BN20052006}.
Both quantities  are related via 
\begin{equation}
J =  C / \tau \quad ,
\label{eq:J}
\end{equation}
where $C$ is a measure for the capacity of the channel.
The mean first passage times for a single replica to cross the system defined by Eq.~(\ref{eq:MET}) in both directions is given by~\cite{NH2007,footnote}
\begin{equation}
\tau = (N+1) \sum_{i=0}^{N-1}{1\over W\left(\beta_i,\beta_{i+1}\right)} \quad,
\label{eq:tau}
\end{equation}
while the channel capacity is simply the number of temperature values, $C=N+1$.
It was shown in Ref.~\onlinecite{NH2007} that 
--- for a particular number of control parameter values ---
 the current is maximized (and therefore the  first passage time minimized) 
 if the flow distribution  is linear in the temperature number. 
This criterion allows an optimization of the temperature distribution~\cite{Trebst2006a,NH2007}. 
We will assume in the following that such an optimized distribution of temperatures has been obtained. In this case, the effective
transition probabilities in Eq.~(\ref{eq:MET}) are constant across the chain of temperatures~\cite{NH2007},
\begin{equation}
W_{opt}(\beta_i,\beta_{i+1}) = const \quad {\rm for} \quad i=0,\ldots,N-1 \quad,
\label{eq:Wopt}
\end{equation}
and this property will be essential in the analysis below.

With increasing number of temperatures, i.e. finer discretization, the transition probabilities approach their maximum $W\to W_0$. Its numerical  value depends on the particular implementation of the 
replica exchange algorithm and the choice of the exchange time scale. 
In this limit the mean first passage time shows the asymptotic behavior
\begin{equation}
\tau\propto N^2 \quad ,
\label{eq:tauN2} 
\end{equation}
i.e. it grows quadratically with number of replicas.
However, for smaller values of $N$ the transition probabilities begin to decrease, leading eventually again to an increase in $\tau$ for small $N$. We are interested in the value of $N$ where $\tau$ is minimal.

This value will depend on the change of the transition probabilities 
 with the control parameter interval, $\left[\beta,\beta'\right]$. 
For the case of  temperature as control parameter, 
this question has been investigated 
in depth~\cite{Predescu2004, Kofke2002, Kofke2004, Kone2005, NH2007}. 
It has been found that the transition probability can be effectively approximated by~\cite{Kofke2004, Kone2005, NH2007,footnoteBBar} 
\begin{equation}
W(\beta,\beta') \approx W_0 f\left(\frac{|\beta-\beta'|}{b}\right) \quad ,
\label{eq:Wf}
\end{equation}
with 
$f(x)$ a monotonically decreasing function obeying $f(0)=1$.
The important quantity here is  $b>0$, denoting the scale of inverse temperatures over which the transition probability decreases. This scale is usually inverse 
to the widths of the  thermal equilibrium energy distributions at $\beta$ and $\beta'$. 
It $decreases$ monotonically with system size and with the extensive heat capcity. In particular, it will be  small near phase transitions. 
More details can be found in Refs.~\onlinecite{Predescu2004,Kofke2002,Kofke2004,Kone2005,NH2007}. 

In the following, we assume  that for a particular system the functional form in Eq.~(\ref{eq:Wf}) is the same over the full temperature range. The only dependence of $W(\beta_i,\beta_{i+1})$ on the  inverse temperature  interval $\left[\beta_i,\beta_{i+1}\right]$ is through the corresponding scale parameter that we denote by $b_{i,i+1}$.
Under this assumption the requirement that --- for the optimal temperature 
distribution --- all effective transition probabilities are constant, Eq.~(\ref{eq:Wopt}), is equivalent to the condition that all 
individual arguments are identical. Hence, 
\begin{equation}
\left|\beta_{i+1}-\beta_i\right|\big/ b_{i,i+1} = const \equiv r \quad 
\label{eq:r}
\end{equation}
holds. Introducing the average  scale $\overline{b}$ 
\begin{equation}
\overline{b} = \frac{1}{N}\sum_{i=0}^{N-1}{b_{i,i+1}}
\label{eq:bBar}
\end{equation}
it can be seen that the constant $r$ has the property
\begin{equation}
r {N\overline{b}} = \beta_{max}-\beta_{min}~.
\label{eq:rbBar}
\end{equation}
This property allows us to introduce the renormalized number of replicas:
\begin{equation}
\nu=\frac{\overline{b}}{\beta_{max}-\beta_{min}} N \quad ,
\label{eq:nu}
\end{equation}
which, in turn, allows us  to cast the mean first passage time  for a replica to cross the system 
into the parameter-free form
\begin{equation}
\tau(\nu) \propto  \nu^2 \Big/ 
f\left(\frac{\displaystyle 1}{\displaystyle \nu}\right) \quad.
\label{eq:taun}
\end{equation}
Minimizing $\nu$ for a particular functional form $f(x)$ of the transition probability decrease 
will give us finally the optimal number of replicas. 

We analyze Eq.~(\ref{eq:taun}) using the following functional forms for $f(x)$,
\begin{equation}
f(x) = \left\{
{\begin{array}{lcr}
\exp(-x)
&  :  & (a) \\
{\rm erfc}\left(\sqrt{\pi/4}x\right)
&  :   & (b) \\
1-x
&  :   & (c) 
\end{array}}
\right.  
\label{eq:fAll}
\end{equation}
In order to ensure they are comparable, we chose $f(0)=1=f'(0)$, i.e. the initial slope is identical for all three forms. 
Form (b) has actually been derived for temperature intervals~\cite{Kofke2004,Kone2005,NH2007,footnoteBBar} and exhibits an
$\exp(-x^2)/x$ tail. It is the one most probable to occur in an actual situation. We have included forms (a) and (c) as worst case scenarios since they cover a wide range of behavior around (b): (a) exhibits a simple exponential decrease, much slower than (b), while (c) exhibits a much faster, linear decrease; note that the latter is valid only for $x<1$.
Figure~1 shows a graphical comparison of the three functions. 

\begin{figure}[t]
\epsfig{file=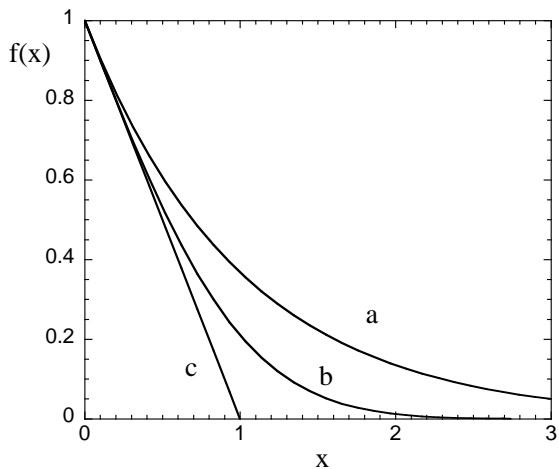,width=7.5cm}
\caption{ Different functional forms for the decay of the transition probability with control parameter distance, compare Eqs.~(\protect\ref{eq:Wf}) and (\protect\ref{eq:fAll}).
\label{fig1}
}
\end{figure}

Minimizing Eq.~(\ref{eq:taun}) gives the optimal value for $\nu$,
\begin{equation}
\nu_{opt} = \left\{
{\begin{array}{lcr}
1/2
&  :  & (a) \\
1.05267
&  :   & (b) \\
3/2 &  :   & (c) 
\end{array}}
\right. 
\label{eq:noptAll}
\end{equation}
These values are all of order one, despite the wide range of functional behavior of the transition probability decrease they describe.
Rewriting Eq.~(\ref{eq:nu}) we obtain our final result for the optimal number of replicas
\begin{equation}
N_{opt}= \nu_{opt}\left(\beta_{max}-\beta_{min}\right) \big/ \overline{b}\quad .
\label{eq:Nopt}
\end{equation}
The ratio of the full temperature range of the simulation to the average scale $\overline{b}$ is the main determining quantity  in that equation. 
In particular, it controls the order of magnitude for $N_{opt}$.
Particular functional forms for the distance dependence of the transition probabilities appear to have less influence since $\nu_{opt}$ is of $O(1)$ only. 
Although the range of values given in Eq.~(\ref{eq:noptAll}) for $\nu_{opt}$ is still covered by factor of three, this is a worst case scenario, and actual values for realistic functional forms will be closer to (\ref{eq:noptAll}b).

\begin{figure}[b]
\epsfig{file=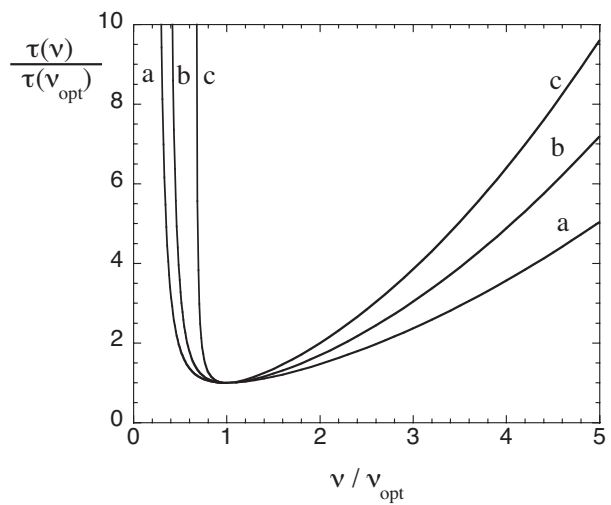,width=9cm}
\caption{ Dependence of relative mean first passage time, Eq~(\ref{eq:taun})
on deviations from the optimal number of replicas
for the different functions of Eq.~(\ref{eq:fAll}).
\label{fig2}
}
\end{figure}

The influence of the different functions, Eq.~(\ref{eq:fAll}), is more important for the form of the minimum of the mean first passage time. Figure~2 shows how the mean first passage time changes when the number of replicas $N$ deviates from the optimal value $N_{opt}$. In order to enable comparison we use the renormalized replica number, Eq.~(\ref{eq:nu}), as variable here. While the minimum is  pronounced for all functional forms, it is steepest for form (c), i.e. the fastest decreasing probability function, and most shallow for form (a).

Before we discuss the consequences of Eq.~(\ref{eq:Nopt}) for protocols to optimize the number of replicas, we need to address some subtleties of the above derivation that we skipped over in favor of a compact derivation: \\
($i$) Equation~(\ref{eq:MET}) is an effective description of the long-time properties of the random walk of replicas in parallel tempering simulations. The transition probabilities for such a long-time description differ from observed acceptance rates, which are, on the replica exchange time scale, short time properties. In particular, it has been observed that for optimized flow the observed acceptance rates are $not$ constant~\cite{Trebst2006a}.
This is due to broken ergodicity~\cite{Palmer} at particular control parameter values.
that gives rise to a hierarchical, tree-like structure for the random walk of replicas~\cite{NH2007}.  
Observed flow and acceptance rates are just projections onto the 1d chain of control parameter values of the more complicated flow processes on the tree. 
However, the possibility of flow optimization shows that for the long-time transition probabilities the property (\ref{eq:Wopt}) holds nevertheless. Such discrepancies between short-time and long-time properties of stochastic processes are well-known~\cite{SSS81,NS8586}.
By using Eq.~(\ref{eq:Wf}) we implicitly assume that those effective transition probabilities exhibit the same qualitative behavior with control parameter difference as it was derived for the short-time transition probabilities.
We feel that this is justified since our qualitative results for $N_{opt}$, Eq.~(\ref{eq:Nopt}), are  independent of the particular functional form. \\
($ii$) $J$ and $\tau$ differ by a factor of $N+1$, see Eq.~(\ref{eq:J}).
The mean first passage time is usually a good estimate for the lowest eigenvalue of the equation system (\ref{eq:MET}), 
i.e. it determines the time scale of equilibration~\cite{SSS81,NS8586},
Since we are interested in fast equilibration, 
 $\tau$ is the more adequate quantity than $J$ to use for comparing systems with different numbers of replicas and to optimize with respect to $N$. \\
($iii$) We have omitted a discussion of the replica exchange time scale. Depending on the frequency of replica exchange moves the time scale of Eq.~(\ref{eq:MET}) may differ by a factor of $N$.
However, our result (\ref{eq:Nopt}) is stable with respect to changes in the power of $N$. Although the particular numerical values change, the value of $\nu_{opt}$ remains a $O(1)$ constant if the exponent in Eq.~(\ref{eq:taun}) changes from two to one ($\nu_{opt}=1,1.6671,2$) or three ($\nu_{opt}=1/3,0.820024,4/3$).

What are the consequences of our results, particularly of Eq.~(\ref{eq:Nopt}), for protocols to optimize the number of replicas? The main result of Eq.~(\ref{eq:Nopt}) is that it identifies, separates, and quantifies  the contribution of various properties of the simulation system to the optimal number of replicas. In particular, it exhibits the quantitative hierarchy of the individual contributions. The above analysis also shows the importance of how  transition probabilities change with the control parameter interval.
To our surprise, for the inverse temperature as control parameter, this complex contribution could be summarized formally into the averaged scale $\overline{b}$, Eq.~(\ref{eq:bBar}). Taking into account the dependence of the scales $b_{i,i+1}$ on the extensive properties of a system~\cite{Kofke2004,Kone2005,NH2007}
suggests that $N_{opt}$ scales with system size $V$ as  $N_{opt}\propto\sqrt{V}$.

We note that the determination of $\overline{b}$ in actual simulations is by no means simple. Since it is defined for the situation of optimal control parameter spacing for a particular number of replicas, such an optimization would have to be performed beforehand. Also, since it describes the behavior of the {\it effective} transition probabilities, see the above discussion, it would have to be determined from the flow distribution together with the actual first passage time, upon slightly varying the discretization.

Instead, 
our analysis suggests that the direct approach to optimize the number of replicas is the most promising one. Comparing first passage times of replicas to cross the simulation system for the optimized discretization is readily possible for different values of $N$. Figure~2, in particular form (b), can then be used as a guideline to extrapolate to $N_{opt}$.

In summary, we have studied the dynamics of parallel tempering simulations. Analyzing these dynamics, we have determined the main factors influencing the optimal number of replicas in such simulations and their quantitative hierarchy. Since the evaluation of the essential term $\overline{b}$, the average scale 
of transition probability decrease, may need costly computations,
we propose to base optimizing the number of replicas on the generic behavior of a replica's first passage time to cross the simulation system given in Fig.~2.
The technique of replica exchange has become the method of choice for the simulation of proteins and other complex systems. The above results add to its understanding, and we believe they will also advance its practical use.

\begin{acknowledgments}
This research was supported by NSF-grant No. CHE-0313618.
\end{acknowledgments}



\begin{thebibliography}{99}
\bibitem{footnoteBBar}
Note that in Refs.~\onlinecite{Kofke2004,Kone2005} the argument used is the ratio $B=\beta'/\beta$; however, since it occurs only in the form $(1-B)/(1+B)$, it is equivalent to the argument in Eq.~(\ref{eq:Wf}).
\end{thebibliography}
\end{document}